# Extending Deep Knowledge Tracing: Inferring Interpretable Knowledge and Predicting Post-System Performance

**Richard SCRUGGS[a]\*, Ryan S. BAKER[a] & Bruce M. MCLAREN[b]**
[a]*Graduate School of Education, University of Pennsylvania, United States*
[b]*Human-Computer Interaction Institute, Carnegie Mellon University, United States*
\*rscr@gse.upenn.edu

**Abstract:** Recent student knowledge modeling algorithms such as Deep Knowledge Tracing (DKT) and Dynamic Key-Value Memory Networks (DKVMN) have been shown to produce accurate predictions of problem correctness within the same learning system. However, these algorithms do not attempt to directly infer student knowledge. In this paper we present an extension to these algorithms to also infer knowledge. We apply this extension to DKT and DKVMN, resulting in knowledge estimates that correlate better with a posttest than knowledge estimates from Bayesian Knowledge Tracing (BKT), an algorithm designed to infer knowledge, and another classic algorithm, Performance Factors Analysis (PFA). We also apply our extension to correctness predictions from BKT and PFA, finding that knowledge estimates produced with it correlate better with the posttest than BKT and PFA's standard knowledge estimates. These findings are significant since the primary aim of education is to prepare students for later experiences outside of the immediate learning activity.

## 1. Introduction

In the last five years, a revolution has been underway in student knowledge modeling. For two decades, a dominant algorithm, Bayesian Knowledge Tracing (BKT; Corbett & Anderson, 1995) was the primary option. Other algorithms, both variants on BKT and – after 14 years – variants on logistic regression (e.g., Pavlik, Cen, & Koedinger's [2009] Performance Factors Analysis [PFA]) and item response theory (Wauters, Desmet, & Van Den Noortgate, 2010) competed with BKT, but the differences in performance between algorithms were small (Gong, Beck, & Heffernan, 2010).

Then, after two decades, Piech et al. (2015) published an exciting new algorithm, Deep Knowledge Tracing (DKT), based on recurrent neural networks, along with initial evidence that its performance at predicting immediate correctness was substantially higher than BKT. Though the difference appears to be somewhat smaller than initially reported, there nonetheless appeared to be a benefit to using DKT instead of BKT (Xiong, Zhao, Van Inwegen, & Beck, 2016; Khajah, Lindsey, & Mozer, 2016). Several papers quickly emerged, proposing extensions and improvements to DKT (e.g., Cheung & Yang, 2017; Yeung & Yeung, 2018; Zhang, Shi, King, & Yeung, 2017; Zhang, Xiong, Zhao, Botelho, & Heffernan, 2017), while other papers explored the utility of additional machine learning methods in predicting problem correctness (e.g., Jiang, Ye, & Zhang, 2018; Lincke, Jansen, Milrad, & Berge, 2019).

However, DKT and its successor algorithms seemed to have two significant limitations relative to earlier approaches. First, DKT as originally implemented produced unstable performance, with oscillating predictions that sometimes went down after producing a correct answer. Yeung and Yeung (2018) proposed a regularization procedure which addresses this limitation. Second, DKT produced only predictions of correctness rather than an estimate of student knowledge on specific human-interpretable skills (see discussion in Pelánek, 2017). The first half of this limitation was addressed by J. Zhang and colleagues (2017), who introduced a skill-item matrix in their approach, Dynamic Key-Value Memory Networks (DKVMN). DKVMN produces predictions of latent knowledge, but with reference to a machine-generated set of skills rather than a human-designed set of skills (the same interpretability limitation is seen in recent approaches that modify BKT to bring its performance into line with DKT – e.g., Khajah et al., 2016).

Indeed, despite DKVMN's stated goal of inferring external knowledge, the initial paper on this algorithm did not attempt to actually predict performance on external measures of knowledge, sticking to the now-dominant paradigm of predicting immediate correctness. What's more, to the best of our knowledge, *none* of the dozens of papers of DKT and its successors have explicitly attempted to measure how well these approaches perform at inferring the knowledge that is carried outside the learning system, through a post-test or other methods, in strong contrast to the early work on BKT, where considerable attention was paid to this goal (e.g., Corbett & Anderson, 1995; Corbett & Bhatnagar, 1997. Also see Pardos, Gowda, Baker, & Heffernan, 2011, for an example including PFA). While other recent papers attempt to tie learning data to skill proficiency (e.g., Wong et al., 2017; Yeung, 2019), their approaches focus on allowing algorithms to discover skills and skill relationships rather than linking back to known and interpretable external skills.

To address this issue, in this paper we attempt to reconnect student knowledge modeling with its initial roots in predicting student knowledge that goes beyond the confines of the learning system. First, we propose a very simple extension that can be applied to DKT, DKVMN, and other algorithms in this family, to enable the algorithms to predict external performance on externally-defined and meaningful skills. This extension consists solely of taking the real-time predicted probability of correctness over all items that a student answered that have been tagged with each external skill, and then calculating the mean of those values for each student, within each skill. To some extent, this follows the "correct first attempt rate" used by Yu et al. (2010) in their KDD Cup winning entry, combining students' performance on all the problems that they attempt. It also captures a student's degree of difficulty in getting to mastery within the system as well as their final state; as Corbett and Bhatnagar (2017) note, final mastery estimates can be incomplete estimates of the knowledge a student carries out of a system when that system has enough practice for most students to reach high within-system proficiency. Although this paper applies this extension solely to predicting performance on an external test, this extension could also be used to report current skill levels to students and teachers in a meaningful fashion (in skill bars, perhaps, as seen in Cognitive Tutors and other platforms [Koedinger & Aleven, 2007]).

Second, we apply this extension to the outputs of DKT and DKVMN, and compare their performance on an external post-test measure of student knowledge to the classic BKT and PFA algorithms. Since it is classic BKT that has extensive evidence for making latent knowledge predictions that are both interpretable and predict post-tests effectively, we use BKT's original formulation rather than modern extensions (i.e., Khajah et al., 2016). Third, we apply this extension to BKT and PFA as well, finding that the same extension improves prediction of post-test performance for these algorithms as well.

In the remainder of this paper, we present each algorithm in greater detail, present the data set that these algorithms will be compared within, discuss results, and then conclude with a discussion of implications and future extensions.

## 2. Algorithms Studied

### 2.1 Deep Knowledge Tracing

Deep knowledge tracing (DKT) uses recurrent neural networks to model student performance learning (Piech et al., 2015). It does not provide estimates of latent student knowledge (unlike BKT), and does not provide estimates of performance for a skill in general (unlike PFA), only predictions of correctness for each actual problem in the data. We implemented DKT using code from Yeung and Yeung (2018), who added extensions to the original method (Piech et al., 2015). The extensions address irregular fluctuations in correctness probabilities as students complete the learning activities and eliminate occasional instances where estimated correctness probabilities either decreased after correct answers or increased after incorrect answers.

In order to generate predictions of external knowledge, we took the probability of correctness over all items that a student answered from each skill, and then calculated the mean of those values for each student, within each skill.[1] These resulting means were then used as knowledge estimates. We refer to these knowledge estimates as coming from mean-DKT.

*2.2  Dynamic Key-Value Memory Networks for Knowledge Tracing*

Dynamic Key-Value Memory Networks (DKVMN) represents states and the relationships within them with two matrices, one for storing internally-derived knowledge components and KC-item mappings and the other for storing the mastery associated with each knowledge component (J. Zhang et al., 2017). While DKVMN produces latent knowledge estimates like BKT, unlike BKT these estimates cannot be straightforwardly mapped back to externally-defined skills, as a new skill-item mapping is distilled bottom-up by DKVMN. Therefore, in order to map DKVMN's estimates back to the posttest, we used the same approach as for DKT: we calculated the mean probability of correctness for each item associated with each skill for each student and used these means as knowledge estimates, referring to those estimates as coming from mean-DKVMN. Code from J. Zhang and colleagues (2017) was used to implement DKVMN.

*2.3  Bayesian Knowledge Tracing*

Bayesian Knowledge Tracing (BKT; Corbett & Anderson, 1995) is an algorithm that infers the probability that students have mastered a skill and the probability they will correctly answer a question which demonstrates that skill. BKT is often thought to differ from other knowledge and performance modeling algorithms in that it explicitly models latent knowledge as well as predicting future correct performance (e.g., Baker, 2019), differentiating between the two with estimates of slip and guess that reflect how performance may not entirely match knowledge. In this study, BKT was implemented using code from Baker et al. (2010), which estimates guess, slip, initial knowledge, and learning transition probabilities for each skill. The parameters were bounded to avoid model degeneracy (Baker, Corbett, & Aleven, 2008), with a floor of 0.01 for all probabilities, a ceiling of 0.3 for guess and slip, and all others having a ceiling of 0.99.

The parameter estimates were applied to the problem data using Excel and the final probability of having learned each skill was recorded for each student. In addition to taking the final probability estimated for each skill, we also calculated knowledge estimates by computing the mean correctness probability for each skill for each student across all of that student's attempted problems, for comparability to the approach used for DKT and DKVMN. We refer to this variation as mean-BKT.

*2.4  Performance Factors Analysis*

Performance Factors Analysis (PFA), pioneered by Pavlik, Cen, and Koedinger (2009), models and predicts student performance using a logistic regression equation that models changes in performance in terms of the number of student successes and failures that have occurred for each skill. PFA estimates the probability of correctness, which is considered as an estimation of learning (Pavlik et al., 2009). In this study, the algorithm was implemented in Excel following the formulas in Pavlik et al. (2009), and using the Excel equation solver to determine optimal parameter estimates. The final learning probability was recorded for each skill for each student.

As with the other algorithms, we also calculated knowledge estimates by computing the mean correctness probability for each skill for each student across all of that student's attempted problems. We refer to this variation as mean-PFA.

## 3. Participants and Data Collection

---

[1] Using DKT, we were unable to calculate valid correctness predictions for 22 problem attempts, out of a total of 70,552 attempts. Those invalid attempts were omitted.

Data from the present study were originally collected for a series of studies, conducted across three semesters, on the effectiveness of erroneous examples on student learning (Richey et al., 2019). The studies aimed to improve students' understanding of decimal numbers and their operations, particularly relating to several common misconceptions held by students (Stacey, Helme, & Steinle, 2001). Participants in the study were sixth-grade students at five urban and suburban schools in the northeast U.S. Data were collected over a six-day period in each study. The materials used in the three studies were the same except that the second and third semester versions of the study had twelve more practice problems than the first. The students received slightly different educational materials depending on whether they were assigned to an erroneous examples group or a more standard problem-solving group. Students in both groups received the same problem content, but erroneous examples problems began by describing a hypothetical student who had answered the problem incorrectly. In both groups, students were then asked to solve a problem (in the case of erroneous examples, this meant finding and correcting the error) and answer an explanatory multiple-choice question about their reasoning. If students responded correctly, they proceeded to the next problem; if they responded incorrectly, they were prompted to answer the incorrect sub-problem(s) with errors again until they got it correct. The materials used did not contain any hints.

A total of 598 students were included in the studies, with 287 students in the erroneous examples group and the remaining 311 in the problem-solving group.

All materials and posttests were delivered through the Tutorshop learning management system, which recorded students' interactions (Aleven, McLaren, & Sewall, 2009). The materials were developed with the Cognitive Tutor Authoring Tools (Aleven et al., 2016). More information about the materials is available in Richey et al. (2019), McLaren, Adams, and Mayer (2015), and Adams et al. (2014). More information about the skills and their relationship to the misconceptions is available in Nguyen, Wang, Stamper, and McLaren (2019).

Students in the study were given 36 (208 students) or 48 (390 students) problems aimed to increase their understanding of decimal numbers. Each of the 36 or 48 problems comprised several subproblems. The problems covered four different skills:
- Ordering decimal numbers by magnitude
- Placing decimal numbers on a number line
- Completing a sequence of decimal numbers
- Adding two decimal numbers

In total, our data set contained 70,552 student attempts at subproblems: 28,908 for ordering decimals, 24,115 for placement on number line, 10,762 for completing the sequence, and 6,767 for decimal addition.

After students completed the problems, their understanding was checked with a 43-item posttest, which tested the four skills. Different numbers of items were used for different skills, in accordance with the number of common misconceptions which were presented for each skill (Richey et al., 2019): 22 items addressed ordering decimals ($M = 0.71$, $S.D. = 0.26$), six addressed placement on a number line ($M = 0.53$, $S.D. = 0.31$), four addressed completing the sequence ($M = 0.59$, $S.D. = 0.28$), and eleven addressed decimal addition ($M = 0.66$, $S.D. = 0.23$).

## 4. Algorithm Application

First, we simplified the students' interaction data, keeping only whether students answered correctly or incorrectly on their first attempt at each problem, in line with common practice in student latent knowledge estimation (Corbett & Anderson, 1995; Pavlik et al., 2009; Piech et al., 2015; J. Zhang et al., 2017). Interaction attempts were then labeled with their associated skill. Next, we trained the set of different student latent knowledge estimation algorithms listed above, using all of the first-interaction data as training data. The implementations of DKT and DKVMN that we used expected separate training and test data sets, however, in this case we used the same data for both sets, since our goal is to understand performance on entirely new external data (posttests) rather than predict future within-system performance. After the algorithms were trained, we derived knowledge estimates for each student using each algorithm. The basic process of training and gathering knowledge estimates was generally similar from algorithm to algorithm, but differed based on how the algorithms treat (or fail to treat) latent knowledge.

## 5. Statistical Comparisons Between Algorithms

After using the four algorithms to produce estimates of latent knowledge for each student and each skill, the estimates were compared. First, we calculated Pearson correlations between each algorithm's knowledge estimates and the posttest scores. As all measurements came from the same population of students, we were able to use a statistical test of the difference in statistical significance between correlations for correlated samples to compare the various correlations to each other (Ferguson, 1976). This test tells us whether one correlation (i.e. one model's ability to predict the post-test) is statistically significantly higher than another correlation (i.e. another model's ability to predict the post-test).

After comparing each combination of algorithms, we performed the Benjamini-Hochberg post hoc control procedure to control for the use of multiple comparisons (Benjamini & Hochberg, 1995; Benjamini & Yekutieli, 2001). This procedure reduces false positives by increasing stringency as more comparisons are performed, maintaining the same false discovery rate regardless of how many statistical tests are conducted.

## 6. Results

Table 1. *Pearson correlations between knowledge estimates and posttest scores*

|  | Ordering Decimals | Placement on Number Line | Complete the Sequence | Decimal Addition |
|---|---|---|---|---|
| mean-DKT | 0.71 | 0.64 | 0.34 | 0.48 |
| mean-DKVMN | 0.72 | 0.62 | 0.35 | 0.56 |
| PFA | 0.28 | 0.33 | 0.10 | 0.26 |
| mean-PFA | 0.69 | 0.64 | 0.36 | 0.49 |
| BKT | 0.44 | 0.43 | 0.28 | 0.49 |
| mean-BKT | 0.65 | 0.52 | 0.28 | 0.44 |

Table 2. *T-scores of correlations between comparisons. * indicates B-H significance at 0.05 level.*

| Ordering Decimals | mean-DKVMN | PFA | mean-PFA | BKT | mean-BKT |
|---|---|---|---|---|---|
| mean-DKT | 1.65 | -14.35* | -1.74 | -10.59* | -3.74* |
| mean-DKVMN |  | -14.67* | -3.22* | -11.86* | -4.26* |
| PFA |  |  | 14.17* | 3.84* | 9.89* |
| mean-PFA |  |  |  | -10.77* | -2.54* |
| BKT |  |  |  |  | 6.43* |

| Placement on Number Line | mean-DKVMN | PFA | mean-PFA | BKT | mean-BKT |
|---|---|---|---|---|---|
| mean-DKT | -1.35 | -9.72* | 0.08 | -8.07* | -6.95* |
| mean-DKVMN |  | -8.85* | 1.26 | -7.72* | -5.31* |
| PFA |  |  | 9.79* | 2.53* | 5.04* |
| mean-PFA |  |  |  | -7.94* | -7.14* |
| BKT |  |  |  |  | 2.93* |

| Complete the Sequence | mean-DKVMN | PFA | mean-PFA | BKT | mean-BKT |
|---|---|---|---|---|---|
| mean-DKT | 0.41 | -5.07* | 1.06 | -1.79 | -2.21* |
| mean-DKVMN | | -4.90* | 0.23 | -2.53* | -2.28* |
| PFA | | | 5.49* | 3.33* | 3.38* |
| mean-PFA | | | | -2.25* | -3.45* |
| BKT | | | | | 0.07 |

| Decimal Addition | mean-DKVMN | PFA | mean-PFA | BKT | mean-BKT |
|---|---|---|---|---|---|
| mean-DKT | 3.16* | -5.89* | 0.76 | 0.32 | -2.08 |
| mean-DKVMN | | -7.46* | -2.53* | -3.13* | -4.35* |
| PFA | | | 6.04* | 5.50* | 4.07* |
| mean-PFA | | | | -0.01 | -3.67* |
| BKT | | | | | -1.57 |

Table 1 shows the correlation between each algorithm's within-tutor knowledge estimates and posttest performance for each skill. Table 2 shows t-scores of the resulting comparisons, with an indication of which tests remained statistically significant after performing the Benjamini-Hochberg control, with FDR (false discovery rate) set to 0.05, equivalent to a p-value of 0.05 for a single test. Results for three skills were broadly similar, with mean-DKT, mean-DKVMN, and mean-PFA producing better estimates than traditional PFA and BKT. Mean-BKT produced estimates that outperformed traditional BKT and PFA in several cases, but generally performed lower than mean-DKT, mean-DKVMN, and mean-PFA.

For Ordering Decimals, mean-DKT (r=0.71) and mean-DKVMN (r=0.72) produced the closest knowledge estimates to the posttest scores. Mean-PFA (r=0.69) produced estimates that were significantly worse than mean-DKVMN, but not significantly different from DKT.

Mean-BKT's (r=0.65) estimates, although close to mean-PFA, were significantly worse than that algorithm, as well as mean-DKVMN and mean-DKT. Estimates from BKT (r=0.44) and PFA (r=0.28) were significantly worse than the other algorithms, with BKT significantly better than PFA.

All algorithms produced worse results on Placement on Number Line, although the order of the groups did not notably diverge from Ordering Decimals. Mean-DKT (r=0.64), mean-PFA (r=0.64), and mean-DKVMN (r=0.62) all produced estimates that did not differ significantly from each other, but beat mean-BKT (r=0.52), BKT (r=0.43), and mean-PFA (r=0.33). Mean-BKT (r=0.52), however, performed better than PFA and BKT.

Complete the Sequence saw all algorithms struggle compared to the first two skills. Mean-PFA (r=0.36), mean-DKVMN (r=0.35), and mean-DKT (r=0.34) performed approximately equally. BKT (r=0.28) was able to produce estimates that were close to the top three and not significantly different from mean-DKT, although its prediction of the post-test was still statistically significantly worse than mean-PFA and mean-DKVMN. Mean-BKT's (r=0.28) estimates did not significantly differ from BKT, but were worse than mean-PFA, mean-DKVMN, and mean-DKT. PFA (r=0.10) performed significantly worse than all other algorithms for this skill.

For Decimal Addition, mean-DKVMN (r= 0.56) achieved significantly better prediction of the post-test than the other algorithms. In turn, mean-PFA (r=0.49), BKT (r=0.49), and mean-DKT (r=0.48) achieved significantly better prediction than PFA (r=0.26). Although BKT's estimates correlated better with the posttest than mean-BKT (r=0.44), that difference was not statistically significant, but mean-PFA produced significantly better estimates than mean-BKT. This finding may seem non-intuitive, since BKT and mean-PFA achieved the same correlation; it is due to there being a higher correlation between mean-PFA and mean-BKT than between BKT and mean-BKT.

## 7. Discussion and Conclusions

Although Deep Knowledge Tracing and Dynamic Key-Value Memory Networks were not designed to produce estimates of latent knowledge for predefined skills, our approach was able to convert performance predictions made by these algorithms into knowledge estimates, which achieved reasonable correlation to student scores on an external posttest. These estimates were more accurate at predicting the external posttest than estimates from Bayesian Knowledge Tracing, which was designed with the aim of estimating the state of students' knowledge. Mean-DKVMN and mean-DKT's estimates were comparable to or perhaps a little better than estimates provided by the classic knowledge modeling algorithm Performance Factors Analysis. In other words, though deep learning-based models might have been thought to mainly capture performance within the system, with a simple adjustment they are also better at inferring the knowledge students carry out of the learning system.

Curiously, PFA only performed comparably to mean-DKT and mean-DKVMN when the same adjustment was made to PFA as was necessary for DKT and DKVMN: averaging estimates across the actual problems, rather than simply taking the final estimate of knowledge for the skill. Explaining this finding may require going back to findings from some of the earliest work in this area. Corbett and Bhatnagar (1997) noted that if mastery learning is used – where a student continues to work within a learning system until the BKT estimate of their knowledge is very high (in that case 0.975) – there is very little variance in the final estimates of student knowledge (as all estimates are above 0.975). However, performance is not always equally high in external post-tests; BKT estimates for students driven to mastery tend to over-estimate post-test performance (Corbett & Anderson, 1995; Corbett & Bhatnagar, 1997). Notably, over-prediction appears to be more characteristic of cases where students had more remedial practice (Corbett & Anderson, 1995). Although the data set used in the current paper did not involve mastery learning, there was a sufficiently large amount of practice in that system (9 to 12 problems per skill for each student) to have caused similar phenomena. For three of the four skills, nearly all final knowledge estimates asymptotically approached either 0 or 1, although students rarely got all posttest items correct or all incorrect. By averaging estimates across problems, we capture student knowledge throughout the learning process rather than apparent knowledge at the end – capturing lower performance on the eventual path to mastery – which appears to be a better estimate of the knowledge students carry out of their learning experience. However, this does not completely explain our results: for Ordering Decimals, no students had final knowledge estimates greater than 0.95 or less than 0.05, but our adjustment still significantly improved the posttest correlations for that skill.

The same adjustment of averaging estimates across actual problems rather than using final knowledge estimates led to better performance for BKT as well as PFA, although not to the same degree. In this paper, the original version of BKT was used. Recent work has suggested that BKT performs better at predicting within-system correctness if several adjustments are made (i.e., Khajah et al., 2016), though still not as well as DKT. It is possible that a version of BKT adjusted in this fashion may perform more comparably to mean-DKT, mean-DKVMN, and PFA for predicting the post-test. However, the very adjustments necessary in Khajah et al. (2016) eliminate some of the benefits – such as interpretable estimates of student knowledge on expert-defined skills – that have made BKT an attractive alternative for practical use.

One of the major arguments in favor of Bayesian Knowledge Tracing has been its interpretable latent estimates – separate from performance. This paper's findings suggest that BKT's latent estimates may not be as useful as thought. BKT does more poorly at estimating an external post-test measure than a reasonable transformation of modern deep learning based algorithms, as well as a more traditional competitor, PFA. Combined with evidence that BKT does more poorly at forecasting time until mastery than PFA (e.g., Slater & Baker, in press), and evidence that classical BKT does more poorly at forecasting future performance within a learning system than DKT or DKVMN (Khajah et al., 2016; J. Zhang et al., 2017), it appears that BKT's use as a primary knowledge modeling algorithm may be coming to an end. With the simple modification to DKT or DKVMN provided here, assessments of specific understandable skills can be provided to teachers and students, one of the core uses of BKT (Koedinger & Aleven, 2007), and these estimates are more predictive of post-test performance than BKT's estimates.

Our findings should not be interpreted as indicating that Bayesian Knowledge Tracing has no use, however. Bayesian Knowledge Tracing still offers the advantage of interpretable parameters, and there are cases – particularly when one wants to understand which skills have low learning rates or high slip rates (e.g., Agarwal, Babel, & Baker, 2018), where BKT may be very useful. In addition, distillations of Bayesian Knowledge Tracing, such as student-level contextual slip, remain useful predictors of long-term outcomes (e.g., San Pedro, Baker, Bowers, & Heffernan, 2013). At this point, however, its shortcomings in predictive accuracy make it harder to justify their use in cases where model structure does not need to be explained.

Of course, no single result is definitive, and more research is needed to establish our findings here as conclusive. This study only investigated data from students' experiences learning decimals in one tutoring system, comparing learning estimates with a single posttest. Our findings, particularly regarding BKT's ability to predict external measures, should be replicated with different student populations and in different domains. However, the results should be encouraging to researchers interested in using DKT, DKVMN, and other cutting-edge knowledge tracing algorithms to infer knowledge, rather than just predicting performance within-system.

There has been considerable work over the last several years to discover which student knowledge model is best at predicting future correctness within intelligent tutoring systems. In Corbett and Anderson's (1995) original vision for student knowledge modeling, as much attention was given to prediction of performance outside the learning system as within it. This seems appropriate, given that the true goal of education is not what students can do during learning, but what they can do beyond and going forward. In this paper, we find that simple enhancements make it possible for recent emerging performance prediction algorithms to also effectively predict knowledge that extends outside the tutoring system. The simple solution provided here will almost certainly fall short of the best that can be done. We hope that in the years to come as much attention will be provided to the problem of predicting long-term and system-external performance as predicting immediate correctness has received recently. Ultimately, the goal of student knowledge modeling should be to infer knowledge, not just predict performance. Happily, it seems like the newest student knowledge algorithms can successfully do this, with only a modest adjustment.